\documentclass[aps,prx,reprint,noshowkeys,nofootinbib, preprintnumbers]{revtex4-1}
\usepackage{amsmath}    % Enhanced math formatting
\usepackage{amssymb}    % Additional symbols
\usepackage{bm}
\usepackage{mathrsfs}
\usepackage{srcltx}
\usepackage{enumitem}
\usepackage{blindtext}
\usepackage{graphicx}   % For figures
\usepackage{newtxmath}  % Times Roman-like math font
\usepackage{braket}
\usepackage{xcolor}
\usepackage{float}
\definecolor{cerulean}{rgb}{0.164706,0.509804,0.635294}
\definecolor{indigo}{rgb}{0.3059,0.3020,0.5412}
\usepackage[colorlinks=true,
    linkcolor=indigo,
    citecolor=cerulean,
    urlcolor=indigo]{hyperref}

\begin{document}
\raggedbottom

\title{Vacuum Decay around Black Holes formed from Collapse}

\author{Giuseppe Rossi}
\email{giusepper@mail.tau.ac.il}
\affiliation{School of Physics and Astronomy, Tel-Aviv University, Tel-Aviv 69978, Israel}

\begin{abstract}
We re-examine the problem of vacuum decay in the presence of spherically symmetric black holes.
Within the semiclassical approximation, we study configurations describing a bubble of true vacuum propagating outside a black hole formed from gravitational collapse.
We find that the saddle point is dominated by a single energy state and that the dependence on the initial conditions in the stellar interior vanishes exponentially fast at late stages of the collapse. Prescriptions are given for implementing the corresponding boundary conditions in the eternal black-hole geometry. 
We find that vacuum decay can only be weakly catalyzed by the black hole, as the suppression exponent attains a minimum at a finite black-hole mass. In the limit of vanishing black-hole mass, the suppression smoothly approaches the flat-space result.
\end{abstract}

\maketitle

\section{Introduction  }
The stability of the vacuum is a fundamental question in particle physics and cosmology, with profound implications for the fate of our universe. In the Standard Model, the shape of the Higgs potential suggests that our current vacuum may be metastable \cite{Sher:1988mj, Isidori:2001bm, Degrassi:2012ry}, allowing for the possibility of a transition to a lower-energy ``true vacuum'' state. It has long been claimed that the presence of a black hole can alter the dynamics of vacuum decay, potentially increasing the decay rate
\cite{Berezin:1987ea,Arnold:1989cq,Berezin:1990qs,Moss:1984zf}. 
These works have employed the instanton method \cite{Coleman:1977py,Callan:1977pt}, which requires solutions to be periodic in Euclidean time. This period is interpreted as the temperature of a thermal bath, which can assist the process of vacuum decay \cite{Affleck:1980ac, Linde:1981zj}.
Applying this method to the black hole case, it is found that the decay rate is controlled by its Hawking temperature, which increases as the black hole emits its radiation. As a consequence, it has been suggested that small black holes could trigger unsuppressed thermal transitions before evaporating, potentially limiting the compatibility of primordial black holes and a metastable Higgs vacuum \cite{Burda:2015yfa,Tetradis:2016vqb,Gregory:2013hja,Kohri:2017ybt, Hayashi:2020ocn, Burda:2015isa, Dai:2019eei} (see also \cite{Strumia:2022jil} for a contrasting assessment). 
The relevance of this result to realistic black holes has been questioned \cite{Gorbunov:2017fhq}, as one would expect that the catalyzing effects of gravity become irrelevant as the black hole becomes smaller, despite its increasing temperature. 
Despite these developments---including qualitative arguments against catalysis \cite{Gorbunov:2017fhq} and the observation that purely thermal saddles do not possess the correct number of negative modes to describe vacuum decay\cite{Briaud:2022few}---a first-principles understanding of this phenomenon in four dimensions is still lacking.\footnote{See \cite{Shkerin:2021zbf, Shkerin:2021rhy} for a two-dimensional model where tunneling in the Unruh state is derived using the complex paths method.} Therefore, a central question is whether vacuum decay around a black hole should be understood as a purely quantum tunneling process or as a thermally assisted transition, and how to compute the corresponding decay rate. 
In this work, we provide a first-principles analysis that addresses these issues. Our approach is to construct semiclassical solutions that originate in the false vacuum prior to gravitational collapse. Then, to show that the result does not depend on the details in the stellar interior, we study solutions that start as true vacuum bubbles excited at fixed energy. We find that dependence on the initial conditions is erased by the large gravitational redshift at the black hole horizon.  
This approach reveals that the standard thermal periodicity prescription---while highly effective in equilibrium settings---does not capture the relevant saddle in a spacetime containing an event horizon formed from collapse. 
We find that the process is dominated by a solution with definite Schwarzschild energy. 
The existence of this critical energy was first noted by Arnold \cite{Arnold:1989cq} in his analysis of the quantum gravity solutions previously proposed by Hiscock \cite{Hiscock:1987hn}. Arnold argued against the physical relevance of this specific value, primarily because no fundamental justification could be provided for selecting a solution that fails to match the Hawking temperature of the black hole. 
Hiscock had proposed such solutions as a zero-temperature idealization, limited only as an approximation for sufficiently large black holes.
We show that this solution is the one relevant for collapse independently of the black hole mass.
Unlike the thermal case, the resulting suppression exponent depends on the black-hole mass in a non-monotonic way: for small black holes the flat-space limit is recovered, the exponent decreases only mildly and attains a minimum at a finite mass, and for sufficiently large black holes it increases indefinitely. From a phenomenological point of view, these results suggest that bounds on primordial black holes inferred from Higgs-vacuum decay should be reassessed, since the strong enhancement expected at small masses does not arise.
The remainder of the paper is organized as follows. 
Section~II introduces the collapse background and the thin-wall effective description of vacuum bubbles, and formulates the semiclassical problem with the relevant boundary conditions.
In Section~III we classify the spherically symmetric saddle-points in the interior region and in the exterior Schwarzschild region, and show that the matching conditions select a unique late-time solution whose exterior evolution is insensitive to the detailed interior collapse history, and compute the exponential suppression.
Section~IV explains how this selection can be phrased directly in the eternal Schwarzschild geometry as a horizon-regularity condition, and clarifies the relation between the Euclidean background metric and our solution. 
Section~V summarizes the main results and discusses future directions.

\section{ Theoretical Framework}
\subsection{Background Geometry and Gravitational Collapse}
\begin{figure}[htb] \centering \includegraphics[width=0.25\textwidth]{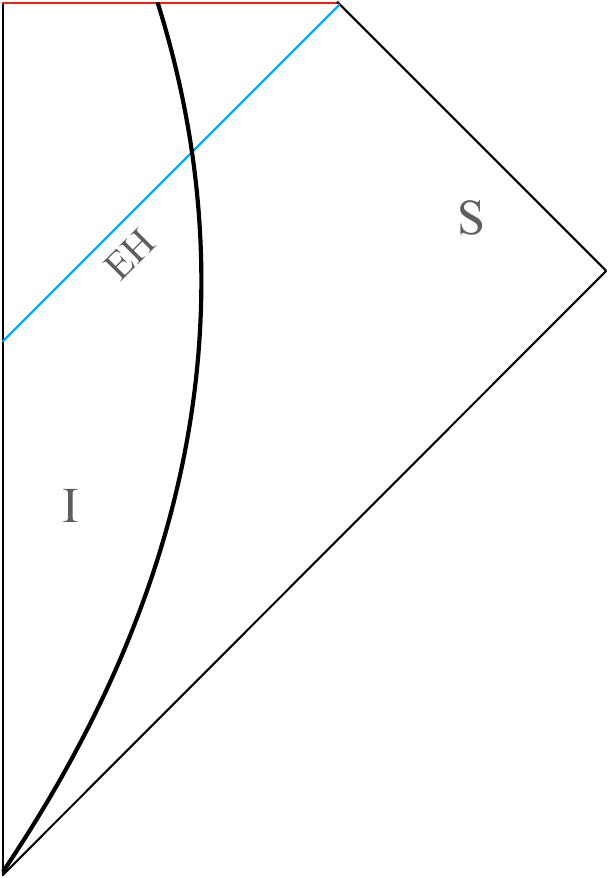} \caption{Penrose diagram of a spherically symmetric spacetime formed from gravitational collapse. The interior region $I$ is matched to an exterior Schwarzschild geometry (S) across the stellar surface (solid line), which follows a timelike trajectory. The event horizon (EH) forms at late times.} \label{fig:penrose-collapse} \end{figure}

We start by outlining the key features of the gravitational background relevant for our analysis. Gravitational collapse to a black hole exhibits certain universal characteristics that will be crucial for what follows, independent of the specific details of the model.
In a spherically symmetric collapse, the stellar surface divides spacetime into two regions: an interior region containing the infalling matter, and an exterior region described by the Schwarzschild geometry (see Fig.~\ref{fig:penrose-collapse}). At early times, the stellar radius lies outside the gravitational radius. As the collapse proceeds, an observer on the stellar surface reaches the gravitational radius in finite proper time. 
However, when described in Schwarzschild time \(t\), the surface asymptotically approaches the horizon and never crosses \(r_s\) at any finite value of \(t\).
At late times, this approach becomes exponentially slow due to the large near-horizon redshift \cite{1967hea3.conf..259T, Misner:1973prb}.
This asymptotic behavior will play a central role in the analysis developed in the following sections.

\paragraph{The Unruh Model of Collapse ----}
To make the general features of gravitational collapse concrete, we now turn to an explicit example, which will serve as our working model throughout this paper.
This model, introduced by Unruh \cite{Unruh:1976db}, consists of a flat interior region matched to an exterior Schwarzschild geometry across a collapsing spherical shell.
The line element takes the form
\begin{equation}
\label{eq:UnruhModel}
ds^2 =
\begin{cases}
 -dt_M^2 + dr^2 + r^2 d\Omega^2, & r < R(t_M), \\[6pt]
 -f(r)\,dt^2 + \dfrac{dr^2}{f(r)} + r^2 d\Omega^2, & r > R(t_M),
\end{cases}
\end{equation}
with \( f(r) = 1 - r_s/r \).
The shell trajectory is specified in terms of the interior Minkowski time \(t_M\) as
\begin{align}
\label{eq:star}
R(t_M) = R_0 - \nu\, t_M\, \theta(t_M),
\end{align}
so that the shell begins collapsing at \(t_M=0\) with constant velocity \(\nu\) and we will take $\nu \simeq 1$ to simplify equations.
We assume \(R_0 > r_s\), ensuring that the shell initially lies outside the Schwarzschild radius.
The shell reaches \(r=r_s\) at the finite Minkowski time
\begin{align}
\label{eq:crossing}
t_{M,\mathrm{coll}} = \frac{R_0 - r_s}{\nu}.
\end{align}
The relation between  \(t_M\) and the exterior Schwarzschild time \(t\) is fixed by requiring continuity of the induced metric across the stellar surface.
One finds
\begin{align}
\label{eq:dtdtM}
\frac{dt}{dt_M}
= &
\left( 1 - \frac{r_s}{R_0} \right)^{-1/2}\,\theta(-t_M)
+ \\ & +
\left[
\frac{R_0 - \nu t_M}{\left( R_0 - r_s - \nu t_M \right)^2}
\left( R_0 - r_s - \nu t_M + r_s \nu^2 \right)
\right]^{1/2}
\theta(t_M) .
\end{align}
As the shell approaches the horizon, this relation simplifies to  
\begin{align}
\frac{dt}{dt_M} \simeq \frac{r_s}{t_{M,\mathrm{coll}} - t_M}. 
\qquad t_M \to t_{M,\mathrm{coll}}^{-}.
\end{align}
Integrating, one finds 
\begin{align}
\label{eq:timeDil}
 t_{M,\mathrm{coll}}  - t_M = A e^{-t/r_s},
\qquad t \to \infty.
\end{align} 
where in our specific model, $A =  (R_0 -r_s) \exp(R_0/r_s -1 ) $. 
This exponential relation between \(t_M\) and \(t\) is not a peculiarity of the Unruh model, but rather a generic consequence of horizon formation when described in Schwarzschild coordinates, and it is a crucial aspect behind Hawking radiation \cite{Unruh:1976db}. 
As anticipated, using this expression in Eqs.~\eqref{eq:star} and \eqref{eq:crossing}, we find that the surface of the star approaches the horizon at an exponentially slow rate
\begin{align}
\label{eq:StellarRadius}
R(t) \simeq r_s +  \nu \, A e^{- t/r_s} \qquad t \to \infty.
\end{align}
We emphasize that the choice of this model is motivated by simplicity: the Minkowski interior possesses a time-translation symmetry that will allow us to solve the bubble dynamics exactly.
At the same time, we will show that the dependence on the details of the interior dynamics disappears exponentially fast as the horizon forms.

\subsection{Vacuum Decay in the thin Wall Approximation}
Vacuum decay proceeds with the nucleation and subsequent expansion of bubbles of the low-energy phase in the false vacuum. 
For concreteness, we will have in mind a field-theoretic model consisting of a single real scalar field $\phi$ with action
\begin{align}
I[\phi] = \int d^4x \,\sqrt{-g}\left[-\frac{1}{2}(\nabla\phi)^2 - V(\phi)\right],
\end{align}
where the potential $V(\phi)$ possesses a metastable false vacuum $\phi_{fv}$ and a lower-energy true vacuum $\phi_{tv}$ separated by a potential barrier. 
In this work, we will assume that the bubble interpolating between the two vacua is spherical and concentric with the black hole. 
When the energy density difference between the two phases is small and the bubble's dimensions significantly exceed its thickness (as measured by the proper distance along geodesics orthogonal to the wall), the system can be analyzed using the thin wall approximation.
In this limit, the only relevant degrees of freedom are the collective coordinates of the interface and the system can be described by a Nambu-Goto action 
\begin{align}
I = - \sigma \int_{\partial V} d^3 \xi \, \sqrt{- h} + \epsilon \int_V d^4 X \, \sqrt{-g}.
\end{align}
Here, $h$ is the determinant of the induced metric on the bubble, $\sigma$ its tension, and $\epsilon$ is the difference in energy density between the two vacua and we have neglected the gravitational back-reaction of the bubble. Explicitly, 
\begin{align}
\epsilon \equiv V(\phi_f)-V(\phi_t), 
\end{align}
and 
\begin{align}
\sigma
= \int_{\phi_{tv}}^{\phi_{fv}} d\phi\,
\sqrt{2\left[V(\phi)-V(\phi_t)\right]} \, .
\end{align} 
Under these assumptions and after gauge fixing, the dynamics is reduced to the analysis of a one-dimensional quantum mechanics system, which can be studied using the semiclassical method \cite{Kobzarev:1974cp}. 
This description is consistent as long as the induced metric on the bubble is non-degenerate $h \neq0$, a condition that we will have to check as we discuss the motion of the bubble.

\subsection{Boundary Conditions and Semiclassical Formulation}
\label{sec:VacPersist}

The transition amplitude between two field configurations $\phi_i(\mathbf{x})$ and $\phi_f(\mathbf{x})$, specified at times $t_i$ and $t_f$, may be written as the path integral
\begin{align}
\label{eq:TransitionAmplitude_Schw}
\mathcal{A}\!\left[\phi_f,t_f;\phi_i,t_i\right]
=
\int_{\substack{\phi(t_i,\mathbf{x})=\phi_i(\mathbf{x})\\ \phi(t_f,\mathbf{x})=\phi_f(\mathbf{x})}}
\!\!\!\mathcal{D}\phi\;
\exp\!\bigl(i I[\phi]\bigr).
\end{align}
The initial time is chosen to be before collapse, and we will consider two classes of initial conditions. The first corresponds to the false vacuum everywhere
\begin{align}
\label{eq:InitialCond}
\phi(t_i,\mathbf{x})=\phi_{\rm fv}.
\end{align}
The second involves excited initial data, in which a true--vacuum bubble is already present inside the stellar interior at $t_i$. The motivation for introducing this second class is to provide a controlled way to test the sensitivity of the late--time decay rate to the preparation of the initial state. As we will show, the suppression exponent at late--times decouples from the detailed initial conditions in the stellar interior.

In the semiclassical approximation, $\mathcal{A}$ is dominated by saddle points compatible with the boundary data. Writing $\bar I$ for the action evaluated on the relevant saddle, one has
\begin{align}
\label{eq:A_semiclassical}
\mathcal{A}\!\left[\phi_f,t_f;\phi_i,t_i\right]
\sim
\exp\!\bigl(i\,\bar I\bigr)\times(\text{prefactor}),
\end{align}
so that an imaginary part of $\bar I$ leads to exponential suppression. 
Because the collapse geometry breaks time translations in Schwarzschild time, the dependence of the transition probability on $t_f$ need not be fixed \emph{a priori}.
However, in the late--time regime relevant for a distant observer, the dominant saddle becomes approximately invariant under $t$--translations.
As a consequence, it possesses a time--translation zero mode (the bounce ``center''), so the one--bounce contribution is proportional to the available time interval.
Summing over configurations with multiple well--separated bounces then exponentiates this linear behavior into the standard decay law, in the same way as in flat space
(see e.g.\ Refs.~\cite{Callan:1977pt,Coleman:1977py}).
We therefore define the decay rate $\Gamma$ through the late--time survival probability of the false vacuum,
\begin{align}
\label{eq:Psurv_def}
P_{\rm surv}(t_f)\; \sim \;\exp\!\bigl(-\Gamma\,t_f\bigr),
\qquad
t_f\to\infty,
\end{align}
equivalently $P_{\rm dec}(t_f)=1-P_{\rm surv}(t_f)$.
In what follows we focus on the leading semiclassical dependence
\begin{align}
\Gamma \;\sim\; e^{-B_S}\, .
\end{align}
where $B_S$ is determined by the imaginary part of the saddle-point action relative to the initial configuration 
\begin{align}
B_S \;=\; 2\,\mathrm{Im}\!\left(\bar I-\bar I_{\rm i}\right).
\end{align}

\section{ Semiclassical Solutions}
This section contains the core semiclassical analysis. 
We will proceed as follows:
\begin{enumerate}[label=\Alph*.]
\item First, we classify solutions in the interior region at early times. We discuss both solutions that correspond to the homogeneous state of false vacuum and solutions describing true vacuum bubbles propagating with arbitrary energy. 
\item Then, we study the possible trajectories that describe propagating bubbles in the exterior region. 
\item These two types of solutions are matched across the surface of the collapsing star.
\item We conclude this section by computing the relevant exponential suppression.
\end{enumerate}

\subsection{Solutions in the Interior Region}
For a spherically symmetric thin wall evolving in the Minkowski interior,
the action is
\begin{align}
\label{eq:IM}
I_M[r]
=
-4\pi\sigma \int dt_M \, r^2 \sqrt{1-\dot r^2}
+
\frac{4\pi}{3}\,\epsilon \int dt_M \, r^3 .
\end{align}

This action is invariant under translations of $t_M$, and this allows us to classify solutions in the interior region according to their conserved energy 
\begin{align}
\label{eq:EnergyM}
E_M
=
4\pi\sigma \, \frac{r^2}{\sqrt{1-\dot r^2}}
-
\frac{4\pi}{3}\,\epsilon\, r^3 .
\end{align}
We now analyze the two types of initial conditions.

\paragraph{False-vacuum initial conditions ----}
To satisfy the initial condition \eqref{eq:InitialCond}, the field must start in the homogeneous false vacuum, which has zero energy 
\begin{align}
E_M = 0
\end{align}

This condition admits two possibilities. 
\\ \medskip
\noindent\emph{(I.a) Homogeneous false vacuum.}
The trivial configuration
\begin{align}
r(t_M)=0, \qquad \dot r(t_M)=0 ,
\end{align}
describes a state in which the interior remains everywhere in the false
vacuum,
\begin{align}
\phi=\phi_{fv}, \qquad r < R(t_M),
\end{align}
for all Minkowski times \(t_M\). 
Here the choice \(r=0\) is purely conventional: any point within the horizon would describe the same homogeneous configuration.
The origin is singled out only because we employ Minkowski coordinates that are concentric with the black hole. 
\medskip
\noindent\emph{(I.b) Expanding bubbles in the interior.}
The other solution, which satisfies 
\begin{align}
\label{eq:Type1}
\dot r^{\,2}
=
1-\frac{r_c^2}{r^2}, 
 \qquad r>0 
\end{align}
where 
\begin{align}
r_c \equiv \frac{3\sigma}{\epsilon}.
\end{align}
describes a bubble of true vacuum expanding in the interior. 
Solving~\eqref{eq:Type1} gives the hyperbolic trajectory \cite{Kobzarev:1974cp, Coleman:1977py}
\begin{align}
\label{eq:hyperbola}
r(t_M)
=
\sqrt{r_c^2+\bigl(t_M-t_0\bigr)^2},
\qquad
r \ge r_c ,
\end{align}
which describes a bubble reaching a minimal radius \(r=r_c\) at \(t_M=t_0\) and subsequently expands. The constant \(t_0\) is a zero mode of the solution and reflects the time-translation invariance of the Minkowski interior region. 
Real Lorentzian solutions exist only for \(r \ge r_c\), so there is no real-time Lorentzian solution evolving continuously from the configuration \(r=0\) to a finite-radius bubble.
Nevertheless, the expanding solution~\eqref{eq:hyperbola} admits a smooth analytic continuation into complex time
\begin{equation}
t_M - t_0 \rightarrow - i t_E ,
\end{equation}
resulting in the Euclidean trajectory
\begin{equation}
\label{eq:EuclSegment}
r(t_E) = \sqrt{r_c^2-t_E^2}, \qquad |t_E| \le r_c .
\end{equation}
This segment interpolates smoothly between \(r=0\) at \(t_E=\pm r_c\) and \(r=r_c\) at \(t_E=0\).
Evaluating the action on the Euclidean segment yields the standard tunneling exponent in flat space \cite{Kobzarev:1974cp, Coleman:1977py}
\begin{align}
B_M
=
2 \int_{0}^{r_c} dr\, \frac{4\pi}{3} r^2 \sqrt{9\sigma^2 -r^2 \epsilon^2 }
=
\frac{27\pi^2\sigma^4}{2\epsilon^3}.
\end{align} 
The Euclidean branch describes tunneling from the false vacuum, while the Lorentzian branch describes the subsequent real-time expansion of the bubble.
Such a solution is valid as long as the Euclidean section remains completely in the Minkowski region, that is to say if $r_c < r_s$. 
In the Minkowski interior, the parameter $t_0$ in~\eqref{eq:hyperbola} is the
time--translation zero mode of the saddle: shifting $t_0$ simply relocates the
nucleation event in $t_M$ without affecting its semiclassical weight.
By contrast, the presence of the stellar surface breaks time--translation
invariance in the collapse geometry.
Interior saddles can propagate into the exterior region only if the Lorentzian
branch intersects the stellar surface before horizon formation.
This requires that, at $t_{M,\mathrm{coll}}$, the bubble radius satisfies
$r(t_{M,\mathrm{coll}})\ge r_s$, which translates into the bound
\begin{equation}
t_0 \;\le\;
t_{M,\mathrm{coll}}-\sqrt{r_s^{\,2}-r_c^{\,2}} .
\label{eq:t0_bound}
\end{equation}

Such configurations do contribute to the transition amplitude and can emerge
outside the stellar surface, continuing their evolution until they encounter a
subsequent classically forbidden region.
However, the associated exponent $B_M$ controls the decay probability per unit
\emph{Minkowski} time.
Since the late--time decay rate measured by a distant observer is defined per
unit Schwarzschild time, this contribution does not generate the exponential
law~\eqref{eq:Psurv_def}.
In particular, the integration over $t_0$ fails to produce a factor
proportional to the available Schwarzschild time, and therefore cannot
exponentiate into a uniform late--time decay rate.

\paragraph{Excited initial conditions---}
We now turn to interior trajectories that are not prepared from the homogeneous
false vacuum, but instead correspond to an already--excited bubble in the
Minkowski region at the initial time $t_i$.  For fixed Minkowski energy $E_M$,
the first integral of motion may be written as
\begin{equation}
\dot r^{\,2}
=
1-
\left(
\frac{4\pi\sigma\, r^2}{E_M+\frac{4\pi}{3}\epsilon r^3}
\right)^{\!2}.
\label{eq:rdot2_excited_short}
\end{equation}
Classical motion is possible at radius $r$ when $\dot r^{\,2}(r)\ge 0$.
Evaluating this condition at the horizon scale $r=r_s$ gives
\begin{equation}
E_M+\frac{4\pi}{3}\,\epsilon r_s^3 \;\ge\; 4\pi\sigma r_s^2 .
\label{eq:allowed_rs}
\end{equation}
If \eqref{eq:allowed_rs} holds, the trajectory can reach $r=r_s$ in real
Minkowski time and may intersect the stellar surface.

\subsection{Solutions in the Exterior Region }
\begin{figure}[htb]
    \centering
    \includegraphics[width=0.5\textwidth]{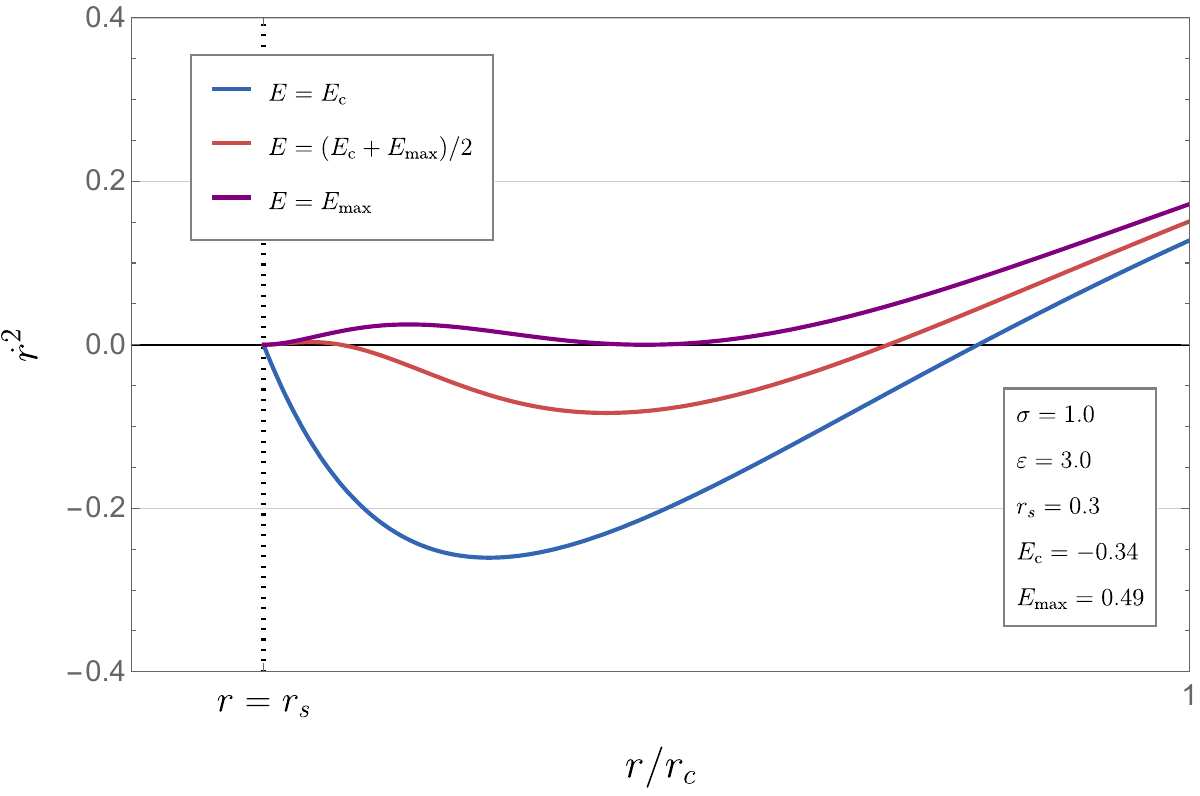}
    \caption{The quantity $\dot{r}^2$ as a function of $r/r_c$ for three different values of the energy $E$. For the critical energy $E = E_c$ the horizon is a turning point, for $E < E_{max}$ there are two turning points, and for $E = E_{max}$ the turning points degenerate into one.}
    \label{fig:rdot_squared}
\end{figure}

We now describe the possible solutions in the exterior region. 
For a true vacuum bubble propagating outside the stellar surface, the Nambu-Goto action in static gauge is 
\begin{align}
\label{eq:SchwarzschildAction}
I_S = - 4 \pi \sigma \int dt \, r^2 \sqrt{f(r) - \frac{\dot{r}^2}{f(r)}} + \frac{4\pi \epsilon}{3} \int  dt \, r^3    \qquad r > R(t). 
\end{align} 
The action is symmetric under time translations, but this symmetry is broken by the boundary. 
This means that for $r > R(t)$ the equations of motion can be reduced to first order, and their energy is determined by matching to the Minkowski region. We will determine the possible values of $E$ in the next section. 
For now, notice that for a fixed $E$, the trajectory satisfies 
\begin{align}
\label{eq:EoMs}
\dot{r}^2 = f(r)^2-\frac{144 \pi ^2 f(r)^3 r^4 \sigma ^2}{\left(3 E + 4 \pi  r^3 \epsilon \right)^2}
\end{align}
or 
\begin{align}
p(r)^2= \frac{9 E^2+24 \pi  E r^3 \epsilon +16 \pi ^2 r^4 \left(r^2 \epsilon ^2-9 f(r)
   \sigma ^2\right)}{9 f(r)^2}.
\end{align}
The motion is classically allowed wherever this expression is positive. We have at most two turning points. For \(E<E_{\max}\equiv \max_{r>r_s}\big[-\frac{4}{3} \pi \epsilon r^{3}+ 4 \pi \sigma r^{2}\sqrt{f(r)}\big]\) there are two simple turning points \(r_1<r_2\) outside the horizon; for \(E=E_{\max}\) they merge into a double turning point and for \(E>E_{\max}\) there are no real turning points. A plot of $\dot{r}^2$ is shown in Fig.~\ref{fig:rdot_squared}, for a variety of values of $E$. 
The most important feature for our discussion is the behavior of the function $p^2(r)$ near the horizon. We notice that there is a critical energy 
\begin{align}
E_c \equiv -\frac{4}{3} \pi \epsilon r_s^{3}
\end{align}
for which the inner turning point sits at the horizon \(r=r_s\) (the outer one remains at some \(r_2>r_s\)). For $E \neq E_c$, the region immediately outside the horizon is always classically allowed. 
Additionally, the function $\dot{r}(t)$ vanishes as $r$ approaches $r_s$, but with a different behavior depending if the value of $E$ is critical or not. For $E = E_c$, we have 
\begin{align}
\label{eq:criticalScaling}
\dot{r} \simeq \pm \frac{i \sigma  \sqrt{r-r_s}}{r_s^{3/2} \epsilon}  
\end{align}
For $E \neq E_c$, we have instead 
\begin{align}
\dot{r} \simeq \pm  \frac{r - r_s}{r_s }.
\end{align}
We note that the difference between these two behaviors has a geometric interpretation, as only for the critical energy the bubble worldvolume remains non-degenerate at the horizon. Indeed, the determinant of the induced metric on the wall behaves as
\begin{align}
E\neq E_c:\qquad
 h \;=\; -2\,(r-r_s)^2\,\sin^2\theta \;+\; O\!\big((r-r_s)^3\big),
\end{align}
\begin{align}
E=E_c:\qquad
h \;=\; -\frac{\sigma^2 r_s^2}{\epsilon^2}\,\sin^2\theta \;+\; O(r-r_s).
\end{align}
Thus, for $E\neq E_c$ the induced metric becomes degenerate, whereas at $E=E_c$ the determinant remains finite. In this sense, the critical branch is the only one for which the thin--wall (Nambu--Goto) description remains internally consistent at the horizon.

% Then in your document body:

\subsection{Matching across the surface of the star}
\paragraph{Matching of Solutions of Type I. a ---- } 
In this first class, the surface of the star remains in the false vacuum at any Minkowski time $t_M$. Since this will be true in particular for the value $t_M = t_{M, \mathrm{coll}}$, we find  
\begin{align}
\label{eq:boundary}
\phi(r < r_s) = \phi_{fv} 
\end{align}
in the strict $t \to \infty$ limit, which is a boundary condition for the eternal black hole consistent with the initial conditions \eqref{eq:InitialCond}. 
From the analysis of the previous section, it is clear that the only possible Schwarzschild trajectory that can satisfy this boundary condition is the one with critical energy, for which the horizon is a turning point. 

The relevant saddle is a ``shell'' configuration: 
the field remains in the false vacuum for $r<r_s$, while a layer of true vacuum occupies a finite interval
$r_s<r<r(t)$ outside the horizon, bounded by the bubble wall at $r=r(t)$. 
In this language the horizon plays the role of an inner endpoint of the wall trajectory, 
and the collapse initial condition selects saddles for which the wall meets $r=r_s$ smoothly. 
This can be seen by working in coordinates that are regular at the horizon. We introduce the proper-distance coordinate $\rho$ defined by
\begin{equation}
r = r_s + \frac{\rho^2}{4 r_s},
\end{equation}
in terms of which the Schwarzschild metric takes the near-horizon Rindler form
\begin{equation}
\label{eq:rindler}
ds^2
=
-\left(\frac{\rho}{2 r_s}\right)^{\!2} dt^2
+ d\rho^2
+ r(\rho)^2 d\Omega^2 .
\end{equation}
To leading order in the near-horizon expansion, the thin-wall action reduces to
\begin{equation}
\label{eq:NearHorizonAction}
I[\rho]
=
-4\pi \sigma r_s^2
\int dt \,
\sqrt{\left(\frac{\rho}{2 r_s}\right)^{\!2} - \dot{\rho}^{\,2}}
\;+\;
\pi \epsilon\, r_s
\int dt \,\rho^2
\;+\;
O(\rho^4)\, .
\end{equation}
This action is invariant under time translations in Schwarzschild time $t$, implying the existence of a conserved Hamiltonian. Direct computation gives
\begin{align}
\label{eq:NearHorizonEnergy}
E
&=
-\frac{4}{3}\,\pi\,\epsilon\, r_s^3
+
\pi \rho^2
\left(
\frac{\sigma}{\sqrt{\left(\frac{\rho}{2 r_s}\right)^{\!2} - \dot{\rho}^{\,2}}}
-\epsilon\, r_s
\right)
\;+\;
O(\rho^4)\, .
\end{align}
The false-vacuum configuration corresponds to $\rho=0$, which again yields the critical energy
$E = E_c$.

Since the saddle lives entirely in the region $r \geq r_s$, Schwarzschild time translations are an exact
symmetry, and this is reflected in the appearance of the collective coordinate
$t_0$: if $r_{\rm cl}(t)$ is a saddle, then so is $r_{\rm cl}(t-t_0)$.
Integrating over $t_0$ produces the usual factor proportional to the available
Schwarzschild time, enabling the exponentiation leading to
Eq.~\eqref{eq:Psurv_def}.

\paragraph{Matching for Trajectories that cross the shell ---- }
Once the bubble crosses the surface of the star, its evolution is governed by the Schwarzschild action~\eqref{eq:SchwarzschildAction}. This is relevant either for trajectories of Type $I.b$ after they have been nucleated in the Minkowski region, or for the excited states trajectories for which the horizon is in the classically allowed region. 
The goal of this section is to determine the Schwarzschild energy of the bubble as a function of the trajectory in the interior. In what follows we present all expressions to leading order in \((1-\nu)\).
Let $r(t)$ describe the worldline of the bubble radius. We denote by $t_p $ the Schwarzschild time at which the crossing occurs and by $dr_p/dt_M $ the Minkowski velocity evaluated at the crossing point. If the bubble wall intersects the surface of the star at radius away from the horizon, the resulting configuration is transient and it is not the object of our computation. 
Thus, we focus on trajectories for which the bubble wall intersects the stellar surface only at late times, when the surface is exponentially close to the Schwarzschild radius.
At this instant, the bubble radius coincides with that of the stellar surface, which can be approximated as 
\begin{align}
r(t_p)
=
R(t_p)
\simeq
r_s + A\, e^{-t_p/r_s}.
\end{align}
and the corresponding Schwarzschild velocity is
\begin{align}
\frac{dr_p}{dt}
&=
\left.\frac{dr_p}{dt_M}\frac{dt_M}{dt}\right|_{t_p}
\simeq \frac{dr_p}{dt_M}  \, \frac{A}{r_s} e^{- t_p/r_s}.
\end{align}
We observe that regardless of the properties of the motion in the Minkowski side, as long as $dr_p/dt_M$ is finite, the Schwarzschild velocity of the bubble approaches zero exponentially fast
\begin{align}
E(t_p)
=
E_c
-
\sqrt{A}\,\bigl(E_M-E_c\bigr)\,e^{-t_p/(2r_s)}
+
O\!\left(e^{-t_p/r_s}\right).
\label{eq:E_late_universal}
\end{align}

This behavior is universal and independent of whether the interior trajectory originates from the Minkowski ground state. This means that even an excited state in the interior will have energy close to critical because of this large gravitational redshift. 
This universality is a direct consequence of the background metric near the horizon, which suppresses sensitivity to the initial conditions and renders the late-time energy insensitive to the interior dynamics, in agreement with the no-hair theorem \cite{ Israel:1967wq, Price:1971fb}.

\medskip

\subsection{Tunneling exponent}
We now turn to the computation of the tunneling exponent of the relevant solution. 
This is given by the reduced action computed on the tunneling trajectory.
For $E=E_c$, the inner turning point coincides with the horizon $r=r_s$, while the outer turning point remains at some $r_2>r_s$. The tunneling exponent is given by
\begin{align}
B_S
&= 2 \int_{r_s}^{r_2} \! dr\, \bigl|p(r)\bigr|
\label{eq:BS_def}\\[4pt]
&= \frac{8\pi}{3}\int_{r_s}^{r_2} \! dr\, \frac{r}{r-r_s}\,
\sqrt{\epsilon^{2}\!\left(r^{3}-r_s^{3}\right)^{2}
      -9\sigma^{2} r^{3}(r-r_s)}
\label{eq:BS_explicit}
\end{align}
As $r_s\to 0$, the outer turning point approaches $r_2=r_c$, reproducing the flat--space result. As $r_s$ increases, $r_2$ decreases, reaching a minimum $r_2\simeq 0.80 r_c$ at $r_s\simeq 0.45 r_c$, before approaching the horizon in the large black hole mass limit.
The tunneling exponent exhibits a mild minimum at
\begin{align}
r_s\simeq 0.24 \, r_c ,
\qquad
B_{\rm min} \simeq 0.87\, B_M ,
\end{align}
showing that black holes can only weakly catalyze vacuum decay. For sufficiently large Schwarzschild radius the exponential suppression increases, indicating that large black holes inhibit the decay process.
\begin{figure}[H]
\centering
\includegraphics[width=0.5\textwidth]{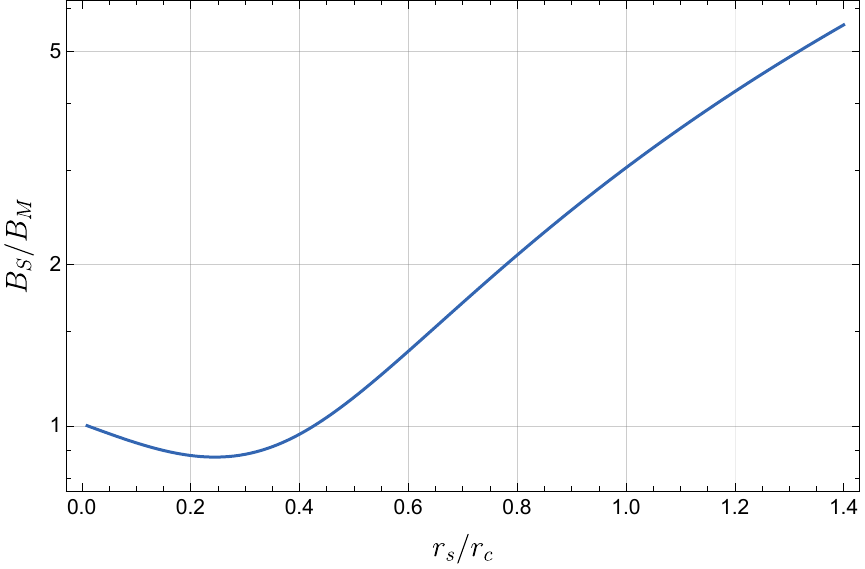}
\caption{Exponential suppression around a Schwarzschild black hole compared to the one in flat space as a function of the black hole gravitational radius.}
\label{fig:veryfinal}
\end{figure}

\section{Euclidean Preparation of the Wavefunction of the Bubble}
We have established that at sufficiently late times—and independently of the interior state—the only semiclassical configuration relevant for the tunneling problem is characterized by a single energy. This observation allows us to formulate the boundary conditions entirely within the eternal black hole geometry, without any reference to the collapsing interior. We show that regularity of the bubble wavefunction at the horizon uniquely selects the critical energy.
Near the horizon, the semiclassical wavefunction takes the form
\begin{align}
\label{eq:WF}
\psi(r)
&\propto
\frac{1}{\sqrt{p(r)}}\exp\left(i \int^{r} p(r'),dr'\right)
\sim
(r-r_s)^{\frac{1}{2} + i\left(E + \frac{4\pi}{3}\epsilon r_s^3\right) r_s},
\end{align}
which exhibits a branch point at $r=r_s$. 
For $ E = E_c$, we have instead 
\begin{align}
\psi(r) \sim (r-r_s)^{1/4} \exp( \pm 8 \pi \sigma r_s^{5/2} \, \sqrt{r-r_s}). 
\end{align}
Although both expressions display a branch point at $r=r_s$, their physical origin is different.
For $E=E_c$, the branch point coincides with a turning point of the radial motion and simply reflects the well--known breakdown of the WKB approximation in the vicinity of turning points \cite{Landau:1991wop}.
In this case, the semiclassical solution may be continued across $r=r_s$ by deforming the integration contour in the complex plane, following the standard method of complex paths.
By contrast, when $E\neq E_c$ the singular behavior occurs away from any turning point and cannot be removed by analytic continuation.
The branch point then signals a genuine pathology of the wavefunction at the horizon, rather than a limitation of the semiclassical approximation.
This structure echoes familiar regularity conditions in Hawking radiation: while the Boulware state \cite{Boulware:1974dm} is singular at the horizon, physically meaningful states such as the Unruh and Hartle–Hawking vacua possess smooth expectation values of observables there \cite{Candelas:1980zt,Birrell:1982ix}.
In the Hawking case, regularity selects different vacua depending on whether smoothness is imposed at the future horizon only or at both past and future horizons. In contrast, for bubble dynamics both criteria lead to the same condition on the energy.
This becomes explicit in ingoing Eddington–Finkelstein coordinates,
\begin{align}
ds^2 = -f(r) dv^2 + 2 dv dr + r^2 d\Omega^2,
\qquad v = t + r_*,
\end{align}
which are smooth at the future horizon and naturally describe a black hole formed from collapse, including the interior region.
In these coordinates, the semiclassical momentum behaves as
\begin{align}
p_v(r)\sim
\begin{cases}
-\dfrac{4 i \pi \sigma r_s^{5/2}}{\sqrt{r-r_s}} + O(1),
& E = E_c, \\[10pt]
\dfrac{2 E r_s + \frac{8}{3}\pi \epsilon r_s^4}{r-r_s} + O(1),
& E \neq E_c,
\end{cases}
\qquad (r\to r_s^+).
\end{align}
which again gives a non-analytic behavior unless the energy is tuned to critical. 

\paragraph{Embedding in the Euclidean Cigar ----}

\begin{figure}[!t]
 \centering
 \includegraphics[width=0.4\textwidth]{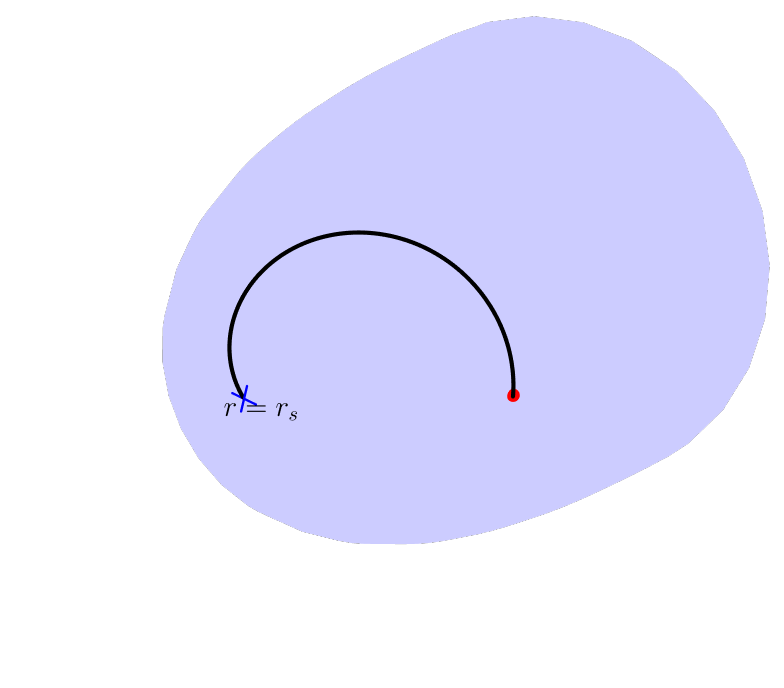}
 \caption{Euclidean Schwarzschild geometry represented as a cigar. 
 The radial direction runs along the axis, while the Euclidean time circle shrinks smoothly to zero size at the horizon. 
 The black curve represents the Euclidean trajectory of the bubble, starting at the horizon (cross) and terminating at the outer turning point (red dot).}
 \label{fig:cigar}
\end{figure}

WKB states such as the one in \eqref{eq:WF} are customarily described using a Euclidean path integral. 
Here we show how our saddle point is smoothly embedded in the Euclidean Schwarzschild geometry. 
For $r>r_s$, the Euclidean Schwarzschild metric is
\begin{align}
ds^2
= \left(1-\frac{r_s}{r}\right)d\tau^2
+ \frac{dr^2}{1-\frac{r_s}{r}}
+ r^2 d\Omega^2 .
\end{align}
Regularity at the horizon requires the identification
\begin{align}
\tau \sim \tau + 4\pi r_s ,
\end{align}
so that near the horizon
\begin{align}
ds^2 \simeq d\rho^2 + \rho^2 d\!\left(\frac{\tau}{2r_s}\right)^2 + r_s^2 d\Omega^2 .
\end{align}
In particular, the $(\rho,\tau)$ section is flat $\mathbb{R}^2$ in polar coordinates. 
The Euclidean Schwarzschild manifold has the shape of a cigar: the $\tau$-circle contracts smoothly to zero size at $\rho=0$, while approaching a finite radius at large $\rho$.
Although $\tau$ is a periodic coordinate, this does \emph{not} mean that the saddle--point configuration must itself be a $\tau$--periodic instanton that wraps the Euclidean time circle. 
Thermal solutions enforce such periodicity by assumption, whereas in our setup the relevant saddle admits a Euclidean representation on the cigar without introducing an additional periodicity constraint \emph{a priori}. 
Geometrically, our solution is an open curve on the cigar: it emanates smoothly from the tip $\rho=0$ (i.e.\ $r=r_s$) and reaches an outer turning point at finite $\rho$. 
A schematic illustration of this trajectory on the Euclidean cigar is shown in Fig.~\ref{fig:cigar}.

\section{Conclusions}
We have revisited vacuum decay in the presence of black holes formed from gravitational collapse. Using the thin-wall approximation and physically motivated boundary conditions, we constructed semiclassical solutions that describe the nucleation of true-vacuum bubbles outside the black hole. Our results show that the decay is not thermally driven.
This might help clarify a tension in the literature. Earlier treatments based on the instanton method have sometimes been interpreted as predicting a strong enhancement of the decay rate for small black holes. In contrast, the boundary conditions appropriate for a spacetime formed by collapse select a different saddle, leading to a qualitatively distinct dependence on the black-hole mass.
Several extensions are worth pursuing. One direction is to compute the fluctuation determinant and verify that the relevant saddle has exactly one negative mode. It would also be useful to check the robustness of the spherical assumption by showing that spherical bubbles give the dominant contribution to the semiclassical decay rate. An additional question is whether the minimum in the suppression persists in the thick-wall regime, which is the physically relevant limit for the Higgs potential. If a similar minimum exists beyond the thin-wall approximation, then the associated suppression could be used to reassess constraints on primordial black holes.

\section{Acknowledgments}

It is a pleasure to acknowledge stimulating discussions with Michael Geller, Andrey Shkerin, Sergey Sibiryakov and in particular Mehrdad Mirbabayi for comments on the manuscript. The work of the author was supported by a scholarship from the Zack's foundation.

\bibliographystyle{apsrev4-2}
\bibliography{biblio}

\end{document}